\documentclass{elsarticle} 

\usepackage{hyperref}
\usepackage{siunitx}
\usepackage{bm}
\usepackage{amsmath}
\usepackage{graphicx}
\usepackage{multirow}
\usepackage{makecell}

\journal{Computer Physics Communications}

\begin{document}

\begin{frontmatter}

\title{Efficient technique for \textit{ab-initio} calculation of 
magnetocrystalline anisotropy energy}


\author[mymainaddress,mysecondaryaddress]{Junfeng~Qiao}

\author[mymainaddress,mysecondaryaddress]{Weisheng~Zhao\corref{mycorrespondingauthor}}
\cortext[mycorrespondingauthor]{Corresponding author}
\ead{weisheng.zhao@buaa.edu.cn}

\address[mymainaddress]{Fert Beijing Institute, BDBC, Beihang University, 
Beijing 100191, China}
\address[mysecondaryaddress]{School of Electronic and Information Engineering, 
Beihang University, Beijing 100191, China}

\begin{abstract}
\textit{Ab-initio} calculation of magnetocrystalline anisotropy energy 
(MCAE) often requires a strict convergence criterion 
and a dense $k$-point mesh to sample the Brillouin zone, 
making its convergence problematic and time-consuming. 
The force theorem for MCAE states 
that MCAE can be calculated by the band energy difference between two 
magnetization directions at a fixed potential. The maximally 
localized Wannier function can be utilized to construct a compact 
Hilbert space of low-lying electron states and interpolate band 
eigenvalues with high precession. We combine the force theorem 
and the Wannier interpolation of eigenvalues together to improve the 
efficiency of MCAE calculations with no loss of accuracy. We use 
a Fe chain, a Fe monolayer and a FeNi alloy 
as examples and demonstrate that the Wannier 
interpolation method for MCAE is able to reduce the computational 
cost significantly and remain accurate simultaneously,  
compared with a direct \textit{ab-initio} calculation on a very dense 
$k$-point mesh. This efficient Wannier interpolation approach makes 
it possible for large-scale and high-throughput MCAE calculations, 
which could benefit the design of spintronics devices. 
\end{abstract}


\end{frontmatter}


\section{Introduction}

In ferromagnetic (FM) materials, the magnetic anisotropy energy (MAE) 
is the difference between magnetizing energies along two directions. 
MAE can be mainly separated into the 
extrinsic part and the intrinsic part. 
The extrinsic shape anisotropy, in which MAE correlates with the 
shape of the FM material, 
originates from the magnetic dipole-dipole interaction \cite{Daalderop1990,Bruno1993}. 
The intrinsic magnetocrystalline anisotropy energy (MCAE)  arises from the collective 
effect of the crystal structure and spin-orbit coupling (SOC). 
The orbital motions of electrons are restricted to specific orientations 
by the crystal field, and the SOC couples the spin degree of freedom and 
the orbital motion. 
Consequently, the energy of the crystal depends on the magnetization 
orientation \cite{Coey2010}.

Usually, the MCAE is on the order of \si{meV} at 
surfaces and interfaces 
\cite{Szunyogh1995Apr,Eisenbach2002Mar,Lazarovits2003Jul}, 
about 3 orders of magnitude 
larger than that of bulk crystals
\cite{Razee1997Oct,Daalderop1990}. 
Apart from the interest in fundamental science, 
the practical applications 
have also stimulated lots of research into 
the understanding and optimization 
of MCAE 
\cite{Wangmengxing2018,Peng2017,Qiao2018,Ouazi2012,Chang2015,Chang2017}. 
One of the applications is the magnetic tunnel junction (MTJ), which 
is the core structure of magnetic random access memory (MRAM) 
\cite{Akerman508,Wangzhaohao2018}. 
MTJ stores one bit of information by the relative magnetization 
orientation of the two FM layers, and the read process is based on 
tunneling magnetoresistance \cite{Mathon2001,Ikeda2008,Zhou2016}. 
The large MCAE induced by the CoFe/MgO interface 
is crucial for the stability of the 
information. 

The MCAE is hard to be captured because of its small magnitude. 
The total energy of a crystal unit cell is on the 
order of \SI{1e3}{eV}, while the MCAE is on the order of 
\SI{1e-3}{eV}, this significant contrast poses serious challenges 
not only to experimental instruments but also to numerical algorithms. 
As a result, calculation of MCAE still remains 
an intriguing problem in the 
\textit{ab-initio} community 
\cite{Daalderop1990a,Daalderop1990,Wang1993,Wang1993a,Blonski2009a}.

In \textit{ab-initio} calculations, plane wave (PW) algorithm 
is widely adopted because it is mathematically simple and in 
principle complete. The PW method provides the same accuracy at 
all points in space, this advantage on one side is appealing but 
on the other side is disappointing since 
it often requires large amounts of plane waves to 
converge and results in bad scaling behavior when system size 
increases. A more efficient approach is always desired. 
Maximally localized Wannier function (MLWF) 
is the Fourier transform of the reciprocal space PW wave function 
under the criterion of maximal localization of Wannier function (WF) 
in real space \cite{Marzari2012}. 
MLWF not only gives a clearer physical insight of chemical 
bondings but also manifests its necessities and accuracies 
in the calculations of many physical properties, 
such as electric polarization \cite{Marzari1998}, 
anomalous hall conductivity \cite{Wang2006} and 
orbital magnetization \cite{Thonhauser2005}, 
etc. \cite{Marzari2012,Ponce2016Dec,Pizzi2014Jan}. 

Here we extend the Wannier interpolation method 
to MCAE calculation for the purpose of improving 
efficiency. Our steps are summarized as follows: 
First, an \textit{ab-initio} calculation is performed on a relatively 
coarse $k$-point mesh (kmesh). Second, MLWFs are extracted from the 
acquired \textit{ab-initio} wave functions. Third, MCAE calculations 
are performed on a much denser kmesh in the Wannier representation. 
Compared to a direct \textit{ab-initio} calculation on the 
dense kmesh, 
this Wannier interpolation approach has equivalent   
accuracy but greatly reduced 
computational cost. The initial coarse kmesh is too sparse 
for MCAE to converge but the extracted WFs meet the criterion of 
``good'' localization. Our Wannier interpolation approach could 
not only reduce the computational cost but also make it possible 
to calculate MCAE for large systems requiring extremal computational 
resources, and calculations requiring exceptionally high accuracy. 

\section{Methods\label{sec:method}}

The Wannier interpolation method for calculating MCAE 
is based on two theories: 
The force theorem (FT) states that MCAE can be calculated by 
the difference between band energies of two magnetization directions; the 
MLWF provides an efficient way for interpolating eigenvalues on 
arbitrary dense kmesh based on coarse kmesh \textit{ab-initio} calculation. 
We first give a brief and self-contained 
introduction to the Wannier interpolation of 
band structure, then show our procedures for combining Wannier 
interpolation with the FT of MCAE. 
We followed the theory and mathematical notations used in 
Yates \textit{et al.} Ref.\cite{Yates2007} and 
Marzari \textit{et al.} Ref.\cite{Marzari2012}. 
The disentanglement of entangled bands is omitted for conciseness 
and detailed theory on MLWFs can be found in 
Ref.\cite{Yates2007} and Ref.\cite{Marzari2012}. 

\subsection{Wannier interpolation}
\subsubsection{Construction of WFs}

In a periodic crystal, the Bloch theorem allows us to write down the 
wave function as 
\begin{equation}
\psi_{n\bm{k}}(\bm{r}) = e^{i \bm{k} \bm{r}} u_{n\bm{k}}(\bm{r}),
\end{equation}
where $\bm{k}$ is the $k$-point vector, $n$ is the band index, 
$\psi_{n\bm{k}}(\bm{r})$ is the Bloch wave function, 
$u_{n\bm{k}}(\bm{r})$ is periodic in real space. 
Inserting Bloch function into the Kohn-Sham equation, we arrive at 
the equation for the periodic part of the Bloch function, 
\begin{equation}
\hat{H}(\bm{k}) u(\bm{k}) = \epsilon_{\bm{k}} u(\bm{k}),
\end{equation}
where 
$\hat{H}(\bm{k})$ is the transformed Hamiltonian 
$\hat{H}(\bm{k}) = e^{-i \bm{k} \bm{r}} \hat{H} e^{i \bm{k} \bm{r}}$, 
$\epsilon_{\bm{k}}$ is the eigenvalue. 
Usually one needs several thousand plane waves 
to expand 
the $u(\bm{k})$ in the PW method 
and the diagonalizations of 
the Hamiltonian matrices having the rank of several thousand are performed on each $k$-point. 
This is why a direct \textit{ab-initio} 
calculation on a dense kmesh is rather time-consuming. 

Generally, Fourier transform enables us to have a different 
representation of functions in reciprocal space, 
and this may be useful for analysis of the problem, e.g. the 
frequency spectrum of audio signals and images. 
If treated properly, the reciprocal space should 
be equivalent to the original space, i.e. they are equivalent 
Hilbert spaces. 
We apply the Fourier transform to the  
Bloch wave function $\psi_{n\bm{k}}(\bm{r})$ which lies in the 
Brillouin Zone (BZ),
\begin{equation}
\label{equ:ftrans}
| \bm{R} n \rangle = \frac{V}{(2 \pi)^3} \int_{\text{BZ}} d \bm{k}
e^{-i \bm{k} \cdot \bm{R}} | \psi_{n \bm{k}} \rangle,
\end{equation}
thus $| \bm{R} n \rangle$ form an orthonormal set and span 
the same Hilbert space as $\psi_{n\bm{k}}(\bm{r})$.
In principle, a smooth function in real space results 
in a localized function in its reciprocal space, and vice versa. 
It is not naturally guaranteed that the simply summed 
Bloch function of Equ. (\ref{equ:ftrans}) results in a 
smooth function $|\bm{R} n \rangle$ in real space. 
Fortunate enough, there is a gauge freedom left in the 
definition of Bloch function, 
\begin{equation}
| \tilde{\psi}_{n\bm{k}} \rangle = e^{i \varphi_n(\bm{k})} 
| \psi_{n\bm{k}} \rangle,
\end{equation}
or equivalently, 
\begin{equation}
| \tilde{u}_{n\bm{k}} \rangle = e^{i \varphi_n(\bm{k})} 
| u_{n\bm{k}} \rangle.
\end{equation}
We can utilize this freedom to construct localized WFs 
in real space, the so-called maximally localized Wannier 
function.

We define the unitary transformation which takes the 
original Bloch function $| u_{n\bm{k}} \rangle$ to 
the smoothed function $| \tilde{u}_{n \bm{k}} \rangle$ 
(from now on written as $| u_{n\bm{q}}^{(W)} \rangle$, 
we use $\bm{q}$ here as to differentiate another kmesh in the 
following Wannier interpolation step) as
\begin{equation}
| u_{n\bm{q}}^{(W)} \rangle = \sum_{m=1}^{M} | 
u_{m\bm{q}} \rangle \mathcal{U}_{mn}(\bm{q}),
\end{equation}
where $M$ is the number of states needs to be considered 
for our targeted physical properties, usually the low-lying 
states below Fermi energy or plus few empty states above. 
We call this unitary transformation as the 
transformation from Bloch gauge to Wannier gauge. 

Thus, the Fourier transform pair between the 
smoothed Bloch functions and the MLWFs are 
\begin{equation}
\begin{split}
| \bm{R} n \rangle & = \frac{1}{N} \sum_{\bm{q}}
e^{-i \bm{q} \cdot \bm{R}} | u_{n \bm{q}}^{(W)} \rangle, \\
& \Updownarrow \\
| u_{n \bm{q}}^{(W)} \rangle & = \sum_{\bm{R}}
e^{i \bm{q} \cdot \bm{R}} | \bm{R} n \rangle,
\end{split}
\end{equation}
where, $N$ is the number of points in BZ.

\subsubsection{Interpolation on arbitrary kmesh}
For the reciprocal space Hamiltonian operator 
$\hat{\bm{H}}(\bm{q})$, we define the $M \times M$ Hamiltonian 
matrix in the Wannier gauge as 
\begin{equation}
H_{nm}^{(W)}(\bm{q}) = \langle u_{n\bm{q}}^{(W)} | \hat{\bm{H}}(\bm{q}) | u_{m\bm{q}}^{(W)} \rangle
= [\mathcal{U}^\dagger (\bm{q}) H(\bm{q}) \mathcal{U}(\bm{q})]_{nm},
\end{equation}
where $H_{nm}(\bm{q}) = \mathcal{E}_{n \bm{q}} \delta_{nm}$, $\delta_{nm}$ is  the 
Kronecker delta function. 
If the $H_{nm}^{(W)}(\bm{q})$ is diagonalized by 
\begin{equation}
U(\bm{q})^\dagger H^{(W)}(\bm{q}) U(\bm{q}) = H^{(H)}(\bm{q}),
\end{equation}
where $H_{nm}^{(H)}(\bm{q}) = \mathcal{E}_{n \bm{q}}^{(H)} \delta_{nm}$, 
then $\mathcal{E}_{n \bm{q}}^{(H)}$ will be identical to the original 
\textit{ab-initio} $\mathcal{E}_{n \bm{q}}$. 

Transforming the Hamiltonian operator from reciprocal space to real space, 
\begin{equation}
H_{nm}^{(W)}(\bm{R}) = \frac{1}{N} \sum_{\bm{q}} e^{-i \bm{q} \cdot \bm{R}}
H_{nm}^{(W)}(\bm{q}),
\end{equation}
and then performing inverse Fourier transform 
\begin{equation}
\label{inv_trans_oper}
H_{nm}^{(W)}(\bm{k}) = \sum_{\bm{R}} e^{i \bm{k} \cdot \bm{R}} 
H_{nm}^{(W)}(\bm{R}),
\end{equation}
we succeed in interpolating the Hamiltonian operator on 
arbitrary $k$-point $\bm{k}$. 
Since the WFs we have chosen are maximally localized, the 
$H_{nm}^{(W)}(\bm{R})$ is expected to be exponentially localized 
in real space, a few $\bm{R}$ are sufficient in the 
sum of Equ. (\ref{inv_trans_oper}). 

The final step is diagonalizing $H_{nm}^{(W)}(\bm{k})$, 
\begin{equation}
U(\bm{k})^\dagger H^{(W)}(\bm{k}) U(\bm{k}) = H^{(H)}(\bm{k}),
\end{equation}
then the acquired eigenvalues on arbitrary $k$-point $\bm{k}$ 
can be used for later extractions of the targeted physical properties. 
We comment here that since $H^{(W)}(\bm{k})$ are of dimensions 
$M \times M$, their diagonalizations are very ``cheap'', compared 
with the diagonalizations of the Hamiltonian matrices in PW method, 
which are on the order of more than $1000 \times 1000$ dimensions.

\begin{figure}
    \includegraphics[width=\columnwidth]{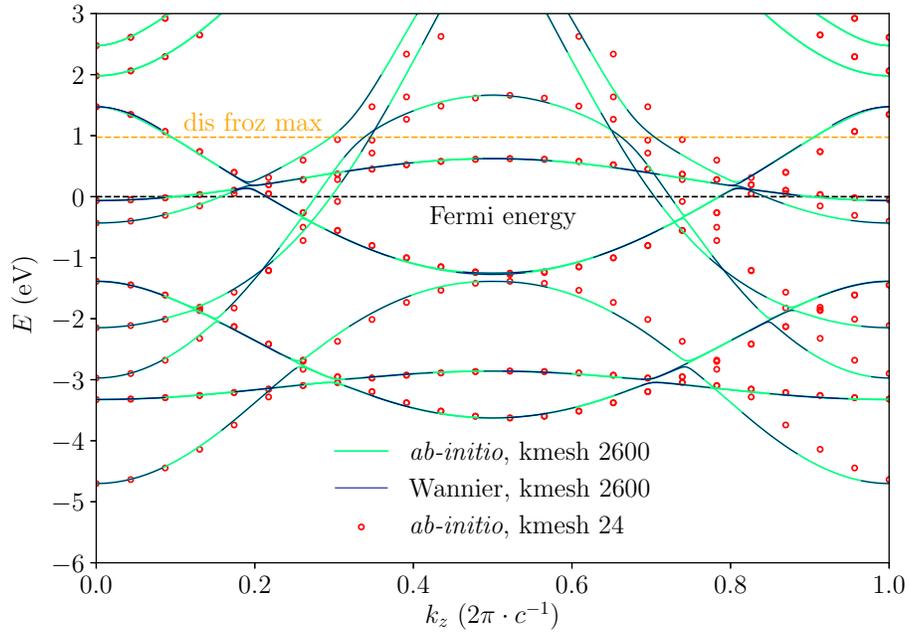}
    \caption{\label{fig:band} 
    	Comparison of Fe chain 
        band structures obtained from 
        a coarse kmesh $1 \times 1 \times 24$ 
        \textit{ab-initio} calculation (red circles), 
        a dense kmesh $1 \times 1 \times 2600$ 
        \textit{ab-initio} calculation (green lines), 
        and the interpolated band structure on the dense 
        kmesh $1 \times 1 \times 2600$
        by MLWFs (blue dashed lines). The MLWFs are 
        constructed from the coarse kmesh 
        $1 \times 1 \times 24$ \textit{ab-initio} calculation. 
        The black dashed 
        horizontal line corresponds to the Fermi energy, 
        the orange dashed horizontal line corresponds to the 
        upper limit of the frozen inner window in the 
        disentanglement process when constructing MLWFs. The 
        $k$-points on the horizontal axis are along the 
        crystallographic $\bm{c}$ axis and are in fractional 
        coordinates. }
\end{figure}

Figure \ref{fig:band} shows the interpolated band structure 
of our later used Fe chain compared with the 
original coarse kmesh \textit{ab-initio} calculation and a dense 
kmesh \textit{ab-initio} calculation. The calculation details 
will be described in section \ref{sec:iron_chain}. The interpolated 
band structure is in excellent agreement with the dense kmesh 
\textit{ab-initio} band, whether it locates at the coarse 
kmesh \textit{ab-initio} $\bm{q}$-points or the interpolated 
$\bm{k}$-points 
between the $\bm{q}$-points. Apparently, the 
coarse kmesh \textit{ab-initio} calculation, from which the 
WFs are constructed, is far from convergence with respect to 
the dense kmesh \textit{ab-initio} calculation. 

\subsection{MCAE}

As mentioned in the introduction, one of the difficulties of 
calculating MCAE is its small magnitude compared with the total 
energy. Another difficulty is its rapid variation in 
$k$-space. As shown in Fig. \ref{fig:struct-mcae_k}(b), 
the $k$-space resolved MCAE, defined by Equ. (\ref{equ:mcae_k}), 
displays sharp peaks and steps along $k_z$ direction. To 
capture these tiny but critical features, a sufficiently 
dense kmesh must be adopted, and slow convergences relative to 
kmesh are often the case in the calculations of MCAE. 

In the \textit{ab-initio} calculation of MCAE, the most 
natural and rigorous method is performing two 
self-consistent (SCF) 
calculations of different magnetization directions, and the 
MCAE should be the difference between the total energies. 
Written down by notations, the MCAE between two crystallographic 
directions can be defined as 
\begin{equation}
\label{equ:mcae_diff}
\text{MCAE} = E^{\ang{90}} - E^{\ang{0}},
\end{equation}
where $E^{\alpha}$ is the total energy of 
magnetization pointing towards the $\theta=\alpha$ direction, $\theta$ is the spherical polar angle 
with respect to the crystallographic $\bm{c}$ axis. 
In our case, we choose $\alpha = \ang{90}$ and $\ang{0}$, 
and the azimuthal angle $\phi$ is kept fixed to $0$ in all 
the calculations. According to this definition, 
$\text{MCAE} > 0$ stands for perpendicular magnetic 
anisotropy (PMA) while $\text{MCAE} < 0$ stands for 
in-plane magnetic anisotropy (IMA).  

However, the cumbersome diagonalizations are rather time-consuming, 
especially when SOC is included, and reaching the 
energy minimal becomes much harder in the SCF iterations. 
According to the FT 
\cite{Razee1997Oct,Wang1993,Daalderop1990,Daalderop1994,Li2013,Li2014}, 
the main contribution to the MCAE originates from the difference of 
band energies between two magnetization directions at fixed potential, 
\begin{align}
\begin{split}
\label{equ:mcae_k}
\text{MCAE}(\bm{k}) & = \sum_{i} f_{i\bm{k}}^{\ang{90}} 
\epsilon_{i\bm{k}}^{\ang{90}}
- \sum_{i^\prime} f_{i^\prime\bm{k}}^{\ang{0}} 
\epsilon_{i^\prime\bm{k}}^{\ang{0}}, 
\\
\text{MCAE} & = \frac{1}{N} \sum_{\bm{k}} \text{MCAE}(\bm{k}),
\end{split}
\end{align}
where $\bm{k}$ is the $k$-point vector, $i$ 
and $i^\prime$ are the 
band indexes of magnetization directions along $\theta=\ang{90}$ 
and $\theta=\ang{0}$, respectively. 
$f_{i\bm{k}}$ is occupation number and  
$\epsilon_{i\bm{k}}$ is the energy of band $i$ 
at $k$-point $\bm{k}$, $N$ is the number of 
$k$-points in the BZ. 

Thus by non-self-consistent (NSCF) 
calculation of only a one-step 
diagonalization, 
the eigenvalues are obtained and the 
MCAE can be calculated 
by Equ. (\ref{equ:mcae_k}). 
The FT is widely adopted and 
its validity has been proved by 
practical calculations 
\cite{Khan2016Oct,Wang1996,Blonski2009a}. 
Good matches between experimental 
and \textit{ab-initio} results are also found in the 
literature \cite{Bairagi2015,Mizukami2011}. 
Another merit of FT is that by Equ. (\ref{equ:mcae_k}) 
some $k$-space resolved analyses can be performed, 
such as the contribution of the quantum well states to 
the oscillation of MCAE \cite{Qiao2018}.  

The practical calculation of MCAE involves two steps. 
First, charge density is acquired self-consistently 
without taking into account SOC. Second, reading the 
SCF charge density, two NSCF 
calculations are performed including SOC, with 
magnetization pointing towards the 
$\theta=\ang{90}$  
and the $\theta=\ang{0}$ directions, respectively.
Finally, MCAE is calculated via Equ. (\ref{equ:mcae_k}).

Before we finish the discussion of 
FT for MCAE, we 
would like to emphasize the importance of Fermi energy 
$E_F$. Since MCAE is such a small quantity that is on the 
order of \si{meV}, a small displacement of $E_F$ 
will propagate into the occupation numbers $f_{i\bm{k}}$, 
and then the MCAE by Equ. (\ref{equ:mcae_k}). The MCAE 
will be significantly modified, even the sign could be 
changed. To accurately calculate MCAE, the small 
difference of Fermi energy between magnetizations 
along $\theta=\ang{90}$ and $\theta=\ang{0}$ must be 
considered. Since the number of electrons $N_{elec}$ 
are the same between two magnetization directions, 
the Fermi energies are calculated separately for 
$\theta=\ang{90}$ and $\theta=\ang{0}$ in 
the Wannier interpolation steps by 
\begin{equation}
N_{elec} = \int^{E_F^{\ang{90}}} n^{\ang{90}}(E) dE 
= \int^{E_F^{\ang{0}}} n^{\ang{0}}(E) dE,
\end{equation}
where $E_F^{\alpha}$ and $n^{\alpha}(E)$ are 
the Fermi energy and density of states for magnetization 
along $\theta = \alpha$, $\alpha = \ang{90}$ or $\ang{0}$. 
Also, the occupation numbers $f_{i\bm{k}}$ are calculated 
separately for the two magnetization directions.  
Finally, MCAE is calculated by Equ. (\ref{equ:mcae_k}). 

In all, Wannier interpolation gives us the ability to 
efficiently interpolate band energies on arbitrary kmesh, 
the FT tells us the MCAE can be calculated by 
the difference of band energies. Combining these two theories 
together, 
MCAE calculations can be carried out on a much denser kmesh 
with no loss of accuracy but greatly reduced 
computational cost. 

We implemented the code for MCAE calculations on the basis 
of {\sc Wannier90} package  \cite{Mostofi2008May,MOSTOFI20142309,Marzari1997,Souza2001}. 
In the next section, we choose a Fe chain as an example 
to demonstrate our method of MCAE calculation and discuss 
some convergence issues in the calculations. 
On one hand, due to the lowered crystal symmetry, the Fe 
chain system is expected to have medium magnitude MCAE; on the 
other hand, this system is small enough that an extremely 
dense kmesh \textit{ab-initio} calculation 
can be performed so that our Wannier interpolation method 
can be directly compared with high accuracy \textit{ab-initio} 
results. After discussing all the critical convergence 
issues in the Wannier interpolation approach, 
we further carry out 
MCAE calculations of Fe monolayer, 
which has MCAE on the order 
of sub \si{meV} and the results can be validated 
against literature. 
Finally, the MCAE of bulk materials 
are on the order of \si{\mu eV} and their 
\textit{ab-initio} calculations are very challenging, 
we show that even in such a tough situation as FeNi 
alloy, our Wannier interpolation approach still faithfully 
recover the converged MCAE from a coarse kmesh calculation, 
and the results well match other \textit{ab-initio} 
calculations and experiments. 
The three examples are respectively 1-dimension, 
2-dimensions, and 3-dimensions, with MCAE varying 
from \si{meV} to \si{\mu eV}. 
Their convergence criteria are increasingly 
tighter, and their densities of 
kmesh increase linearly, quadratically and 
cubically when testing the MCAE convergence. 
These examples are good test grounds 
for verifying Wannier interpolation 
approach of MCAE.

\section{Fe chain}

\subsection{\label{sec:iron_chain}\textit{ab-initio} calculation and Wannierization}
\begin{figure}
    \includegraphics[width=\columnwidth]{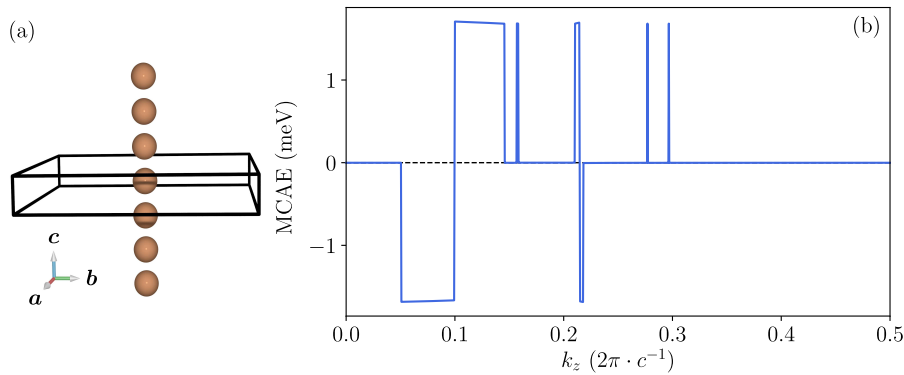}
    \caption{\label{fig:struct-mcae_k}  
        (a) Crystal structure of the Fe chain. 
        The chain is along the crystallographic $\bm{c}$ axis. 
        The cuboid represents the unit cell. 
        (b) $k$-resolved Fe chain MCAE along $k_z$ axis, 
        defined by Equ. (\ref{equ:mcae_k}). Only half of the 
        $k$-space is shown.}
\end{figure}

The \textit{ab-initio} calculations were performed 
using the 
{\sc Quantum ESPRESSO} (QE) package based on the projector-augmented wave (PAW) 
method and a plane wave basis set \cite{Giannozzi2009,Giannozzi2017Oct}. 
The exchange and correlation terms were described using 
generalized gradient approximation (GGA) in the scheme of 
Perdew-Burke-Ernzerhof (PBE) parameterization, as implemented in the 
{\sc pslibrary} version 0.3.1 \cite{Corso2014}. 
We used a wave function cutoff of \SI{90}{Ry} and electron density 
cutoff of \SI{1080}{Ry}. A Marzari-Vanderbilt cold 
smearing \cite{PhysRevLett.82.3296} of width \SI{0.002}{eV} 
was adopted. Convergence relative to smearing width and kmesh will 
be detailedly discussed in the subsequent 
paragraph. Since the MCAE of the Fe chain is on the order 
of \si{meV}, the energy convergence criteria of all the 
calculations were set as \SI{1.0e-8}{Ry}. We kept a 
\SI{15}{\angstrom} vacuum space in the xy plane 
to eliminate interactions between periodic images. 
Since enough vacuum space was left in the xy plane, 
we set the Monkhorst-Pack $k$-point mesh to $1 \times 1 \times 24$ 
and confirmed that increasing it to $2 \times 2 \times 24$ or more 
had negligible impacts on MCAE. The unit cell contains one 
Fe atom [shown in Fig. \ref{fig:struct-mcae_k}(a)] and the 
lattice constant along $\bm{c}$ axis 
(i.e. the atomic spacing along the Fe chain) was set as 
\SI{2.2546}{\angstrom}, which was acquired by relaxation until 
the force acting on the Fe 
atom was less than \SI{1e-2}{Ry/Bohr}. 

To transform Bloch functions into MLWFs, 
first, the overlap matrices 
$M_{mn}^{\bm{k},\bm{b}}$ and the projection matrices $A_{mn}^{\bm{k}}$ 
are extracted from the Bloch functions \cite{Marzari2012}. Then 
after disentanglement and Wannierization the MLWFs are constructed. 
We used \num{20} spinor WFs having the form of 
$sp3d2$, $dxy$, $dxz$, $dyz$, and $s$-like Gaussians. 
The disentanglement outer energy window was set as 
[\num{-100.0},\num{10.0}]\si{eV}, and the inner window was set as 
[\num{-100.0},\num{-3.41}]\si{eV}, which is slightly higher 
than the Fermi energy \SI{-4.38}{eV}. The spreads of WFs signify 
the quality of localization, as defined by \cite{Marzari2012}
\begin{equation}
\Omega = \sum_n [\langle \bm{0} n | r^2 | \bm{0} n \rangle 
- \langle \bm{0} n | \bm{r} | \bm{0} n \rangle^2],
\end{equation} 
where $| \bm{0} n \rangle$ are the WFs in the home unit cell. 
The convergence threshold of spread for both the 
disentanglement and Wannierization were 
set as \SI{1e-10}{\angstrom^2}. Apart from \num{2} WFs having 
the spread $\Omega$ around \SI{2.0}{\angstrom^2}, the 
spreads of the remaining \num{18} WFs were smaller than 
\SI{1.0}{\angstrom^2}, mostly around \SI{0.2}{\angstrom^2}. 
The total spread of WFs was around \SI{10.0}{\angstrom^2} and 
the convergence issues will be discussed in the next section.

\subsection{Convergence issues of Wannier interpolation}

Three main parameters should be tested to reach a reliable 
result of MCAE: the number of coarse \textit{ab-initio} 
kmesh $n_k^{ab}$ for the construction of MLWFs, the 
number of Wannier interpolation kmesh $n_k^{wan}$ for 
the interpolation of MCAE from MLWFs, and the 
smearing width $\sigma$ that determines the fictitious 
smearing contributions to the total energies.

Note since the kmesh in the xy plane is always set as 
$1 \times 1$, we will use the number of $k$-points along 
z-axis 
$n_k$ as the shorthand of the kmesh $1 \times 1 \times n_k$. 
We differentiate the coarse \textit{ab-initio} kmesh and the 
dense Wannier interpolation kmesh by the notation $n_k^{ab}$ 
and $n_k^{wan}$. 

To make a clearer separation among different tests, we 
obey the notation that for a function $f(x;y)$, the 
$x$ before the semicolon is the variable while the 
$y$ after the semicolon is the parameter which is held 
fixed. For example, $\text{MCAE}(n_k^{ab};n_k^{wan})$ 
represents the variation of MCAE relative to the 
coarse \textit{ab-initio} kmesh $n_k^{ab}$ 
when the Wannier interpolation kmesh $n_k^{wan}$ is 
held fixed. 

Next, we test the convergence of the spread relative to 
the number of coarse \textit{ab-initio} kmesh 
$\Omega(n_k^{ab})$, the convergence of MCAE relative 
to the the number of coarse \textit{ab-initio} kmesh 
$\text{MCAE}(n_k^{ab};n_k^{wan},\sigma)$, 
the convergence of MCAE relative to the number of 
Wannier interpolation kmesh 
$\text{MCAE}(n_k^{wan};n_k^{ab},\sigma)$, 
the convergence of MCAE relative to the smearing width 
$\text{MCAE}(\sigma;n_k^{ab},n_k^{wan})$ and 
the minimum number of Wannier interpolation kmesh 
needed relative to smearing width 
$n_k^{wan}(\sigma)$. 

Before carrying out quantitative analyses, we introduce 
several definitions to avoid ambiguities. 
\begin{enumerate}
\item The \emph{convergence indicator} $\delta f(x)$ 
of a function $f(x)$ is defined as 
\begin{equation}
\label{equ:delta_f}
\delta f(x) = | \max_{y \ge x} f(y) - 
\min_{y \ge x} f(y) |,
\end{equation}
where $|\ldots|$ means taking the absolute value. 
\item Let $\epsilon$ be a positive real number, 
we say that the function $f(x)$ is \emph{converged 
at} $x_0$ under the threshold $\epsilon$ 
if $\delta f(x) < \epsilon$ is satisfied for any 
$x \ge x_0$.
\item The \emph{converged value} $\bar{f}$ is defined 
as 
\begin{equation}
\bar{f} = \sum_{\{x|\delta f(x) < \epsilon\}} f(x),
\end{equation}
i.e. if $f(x)$ is converged at $x_0$, then 
$\bar{f} = \sum_{x\ge x_0} f(x)$. 
\item The \emph{deviation} of a function $f(x)$ is 
defined as
\begin{eqnarray}
\Delta f(x) = f(x) - \bar{f}
\end{eqnarray}
\end{enumerate}

\subsubsection{Spread of WFs}

The construction of WFs relies on a uniform 
\textit{ab-initio} kmesh, we call it as coarse 
\textit{ab-initio} kmesh, 
since this kmesh does not need to be dense enough for the MCAE 
to be converged but is sufficient once 
``good'' localization 
of WFs are reached. 

\begin{figure}
    \includegraphics[width=\columnwidth]{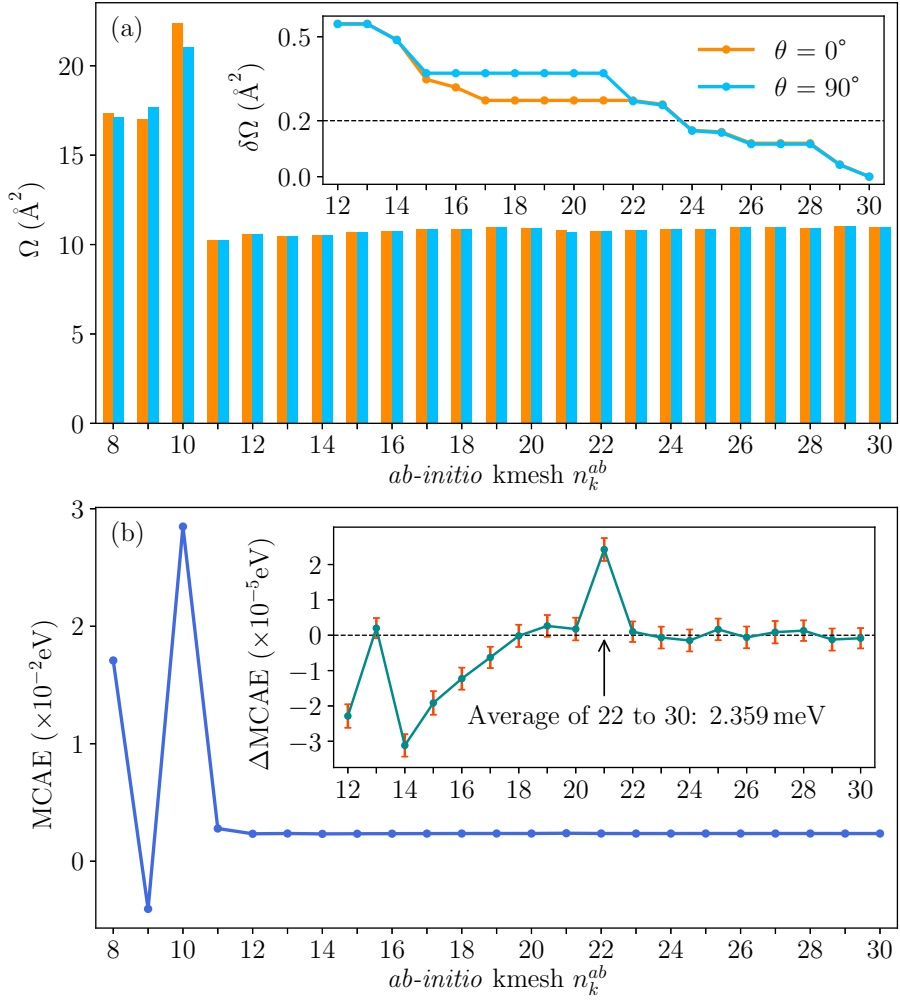}
    \caption{\label{fig:conv_ab} For Fe chain: 
    	(a) Spread of WFs relative to the 
        density of \textit{ab-initio} kmesh, 
        the inset shows the convergence trend $\delta\Omega(n_k^{ab})$ 
        of the spread defined by Equ. (\ref{equ:delta_omega}). 
        Blue color and orange color are the calculations of magnetization direction 
        perpendicular to and along the Fe chain, respectively. 
        (b) The MCAE relative to the density of \textit{ab-initio} kmesh. 
        The inset shows the deviation of MCAE when approaching convergence, defined 
        by Equ. (\ref{equ:Delta_mcae_ab}). The red error bars in the inset 
        represent the convergence trend 
        $\delta\text{MCAE}(n_k^{wan};n_k^{ab})$ 
        relative to $n_k^{wan}$ as defined 
        by Equ. (\ref{equ:delta_mcae_wan}). 
    The small error bars indicate that the anomaly at 
    $n_k^{ab} = 21$ is caused by the \textit{ab-initio} 
    kmesh alone and this means \textit{ab-initio} kmesh 
    plays a critical role in Wannier interpolation of 
    MCAE.} 
\end{figure}

Here we show the relation of WFs' spread and the density of 
\textit{ab-initio} kmesh in Fig. \ref{fig:conv_ab}(a). 
In the MCAE calculations, the ``good'' 
criterion for the localization of WFs is ultimately determined 
by the convergence of the MCAE. We mention it 
in advance that a 
$1 \times 1 \times 2600$ interpolation kmesh 
and a smearing width of \SI{0.0012}{eV} are sufficient 
for MCAE to be converged on the order of \SI{1e-6}{eV}, 
and the converged MCAE is $2.358 \pm 0.001$\si{meV} with 
Fe magnetic moment parallel to the chain. The 
subsequent paragraphs will detailedly discuss the convergence 
issues related to interpolation kmesh and smearing width. We use 
this conclusion here to exclude the influence of the 
Wannier interpolation kmesh and the 
smearing width when discussing 
the convergence of the WFs' spread and the MCAE 
relative to the density of the \textit{ab-initio} kmesh. 

We use the indicator 
$\delta\Omega(n_k^{ab})$ to represent the convergence 
trend of the spread $\Omega$ calculated on \textit{ab-initio} 
kmesh $n_k^{ab}$
\begin{equation}
\label{equ:delta_omega}
\delta\Omega(n_k^{ab}) = | \max_{i \ge n_k^{ab}}\Omega(i) - 
\min_{i \ge n_k^{ab}}\Omega(i) |.
\end{equation}
As shown in the 
inset of Fig. \ref{fig:conv_ab}(a), when $n_k^{ab} \ge 24$, 
the $\delta\Omega$ are less than \SI{0.2}{\angstrom^2}. 
We use $\Delta \text{MCAE}(n_k^{ab})$ to represent the deviation 
of $\text{MCAE}(n_k^{ab})$ relative to the converged 
$\overline{\text{MCAE}}$. In this case, they are defined as
\begin{align}
\begin{split}
\label{equ:Delta_mcae_ab}
\Delta \text{MCAE}(n_k^{ab}) & = \text{MCAE}(n_k^{ab}) 
- \overline{\text{MCAE}}, 
\\
\overline{\text{MCAE}} & = \frac{1}{9} \sum_{n_k^{ab} = 22}^{30} 
\text{MCAE}(n_k^{ab}),
\end{split}
\end{align}
where $n_k^{ab}$ is the \textit{ab-initio} kmesh used for the 
construction of WFs, $\overline{\text{MCAE}}$ is the mean 
value of MCAE calculated with \textit{ab-initio} kmesh 
from $22$ to $30$ and these are the converged value 
since the variation between these \num{9} values are 
on the order of \SI{1e-6}{eV}, as shown in the 
inset of Fig. \ref{fig:conv_ab}(b). 

An interesting anomaly appears in the inset of 
Fig. \ref{fig:conv_ab}(b) at $n_k^{ab} = 21$. 
This may be caused by the insufficiencies of 
coarse \textit{ab-initio} 
kmesh or the Wannier interpolation kmesh. 
To separate the influence of these two kmeshes, 
we show the convergence indicator 
$\delta\text{MCAE}(n_k^{wan};n_k^{ab})$ 
by the red error bar in the inset of 
Fig. \ref{fig:conv_ab}(b). The 
$\delta\text{MCAE}(n_k^{wan};n_k^{ab})$ 
is defined as
\begin{equation}
\label{equ:delta_mcae_wan}
\delta\text{MCAE}(n_k^{wan};n_k^{ab}) =  
| \max_{i \ge n_k^{wan}}\text{MCAE}(i;n_k^{ab}) 
 - \min_{i \ge n_k^{wan}}\text{MCAE}(i;n_k^{ab}) |,
\end{equation}
where $n_k^{wan}$ is in the range 
of \numrange{2500}{2700}. 
At each specific $n_k^{ab}$, 
$\delta\text{MCAE}(n_k^{wan};n_k^{ab})$ only varies less 
than \SI{3e-6}{eV} when $n_k^{wan}$ changes from 
\numrange{2500}{2700}, much smaller 
than the anomaly. 
This indicates that the fixed Wannier 
interpolation kmesh of 
$n_k^{wan} = 2600$ is high enough and the 
anomaly at $n_k^{ab} = 21$ is caused by the \textit{ab-initio} kmesh alone. 
This means to get a very accurate MCAE, 
a sufficiently dense \textit{ab-initio} kmesh 
is needed for constructing high-quality MLWFs. 
However, compared with the \textit{ab-initio} 
kmesh $n_k^{ab} = 2600$ used for a direct 
\textit{ab-initio} computation of MCAE, this 
kmesh is much sparser and only use 
\SI{0.8}{\percent} of the number of $k$-points.

For $n_{k}^{ab} \ge 22$, the total spread of WFs converges 
to approximately \SI{0.2}{\angstrom^2}, at the same time 
the MCAE converge on the order of \SI{1e-6}{eV} as shown 
in the inset of Fig. \ref{fig:conv_ab}(b). 
As in many other cases, the interpolated band structure 
agrees very well with high-density kmesh \textit{ab-initio} 
calculation \cite{Marzari2012}. While the band structure of 
the coarse \textit{ab-initio} kmesh, which is used for the 
construction of MLWFs, deviates largely from the ``true'' 
dense kmesh band structure needed for MCAE to converge. 
The comparison of band structures is shown in Fig. \ref{fig:band}.

\subsubsection{Interpolation kmesh}
The former paragraphs consider the convergence of MCAE with respect 
to \textit{ab-initio} kmesh, which should be dense enough for 
constructing MLWFs. From now on, we fix the $n_k^{ab}$ to $24$. 
Once high-quality MLWFs are constructed, 
the next step for reaching converged MCAE is the Wannier 
interpolation. This is where the computational cost 
is significantly reduced. 

\begin{figure}
    \includegraphics[width=\columnwidth]{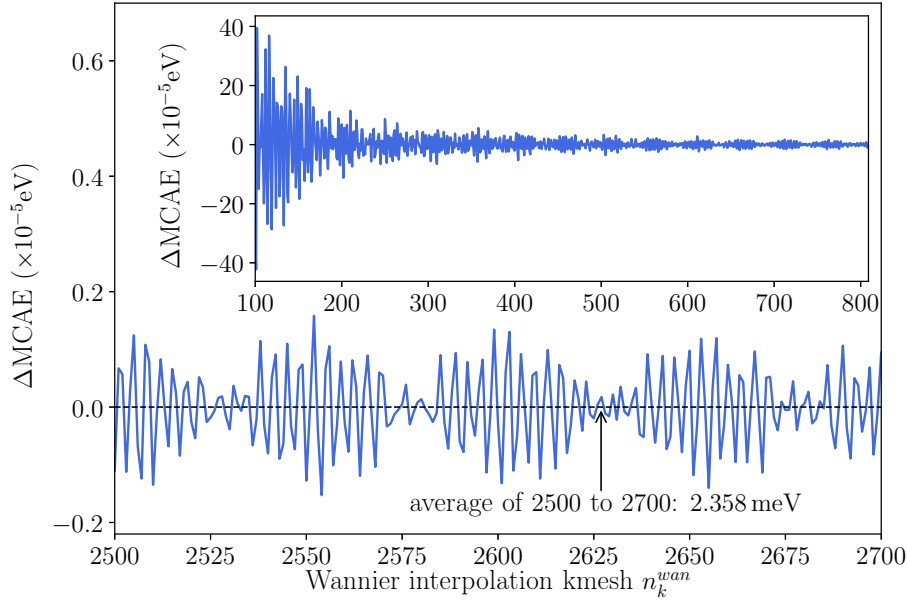}
    \caption{\label{fig:conv_wan} 
        The Fe chain MCAE deviation relative to the 
        converged value when Wannier 
        interpolation kmesh $n_k^{wan}$ varies, 
        defined by Equ. (\ref{equ:Delta_mcae_wan}). 
        The inset is the 
        overview with $n_k^{wan}$ from \numrange{100}{800}.}
\end{figure}

In such circumstance, the $\Delta\text{MCAE}$ is defined as 
\begin{align}
\begin{split}
\label{equ:Delta_mcae_wan}
\Delta\text{MCAE}(n_k^{wan}) & = \text{MCAE}(n_k^{wan}) 
- \overline{\text{MCAE}}, 
\\
\overline{\text{MCAE}} & = \frac{1}{201} \sum_{n_k^{wan} = 2500}^{2700} 
\text{MCAE}(n_k^{wan}),
\end{split}
\end{align}
and this is shown in Fig. \ref{fig:conv_wan}. 

The inset of Fig. \ref{fig:conv_wan} shows the overview 
of the $\Delta\text{MCAE}(n_k^{wan})$ for 
$n_k^{wan}$ from \numrange{100}{800}. 
Apparently, $n_k^{wan}$ on the order of \num{100} is far 
from convergence. For $n_k^{wan}$ around \num{2600}, the 
MCAE converge under \SI{4e-6}{eV}. 

\subsubsection{Smearing contribution}
Smearing is very important for the convergence of 
\textit{ab-initio} calculations. 
Large smearing is beneficial for fast convergence, but 
this may introduce a fictitious and un-negligible contribution to the 
total energy. Too small smearing may slow down the convergence 
significantly, and often one needs a much denser kmesh to converge. 
This problem can be largely alleviated because the Wannier 
interpolation is very ``cheap''\textemdash 
as we mentioned in the Sec.\ref{sec:method} it only involves 
diagonalizations of $M \times M$ matrices. One could 
easily increase kmesh density and reduce smearing width, 
or even not resorting to smearing. 

\begin{figure}
    \includegraphics[width=\columnwidth]{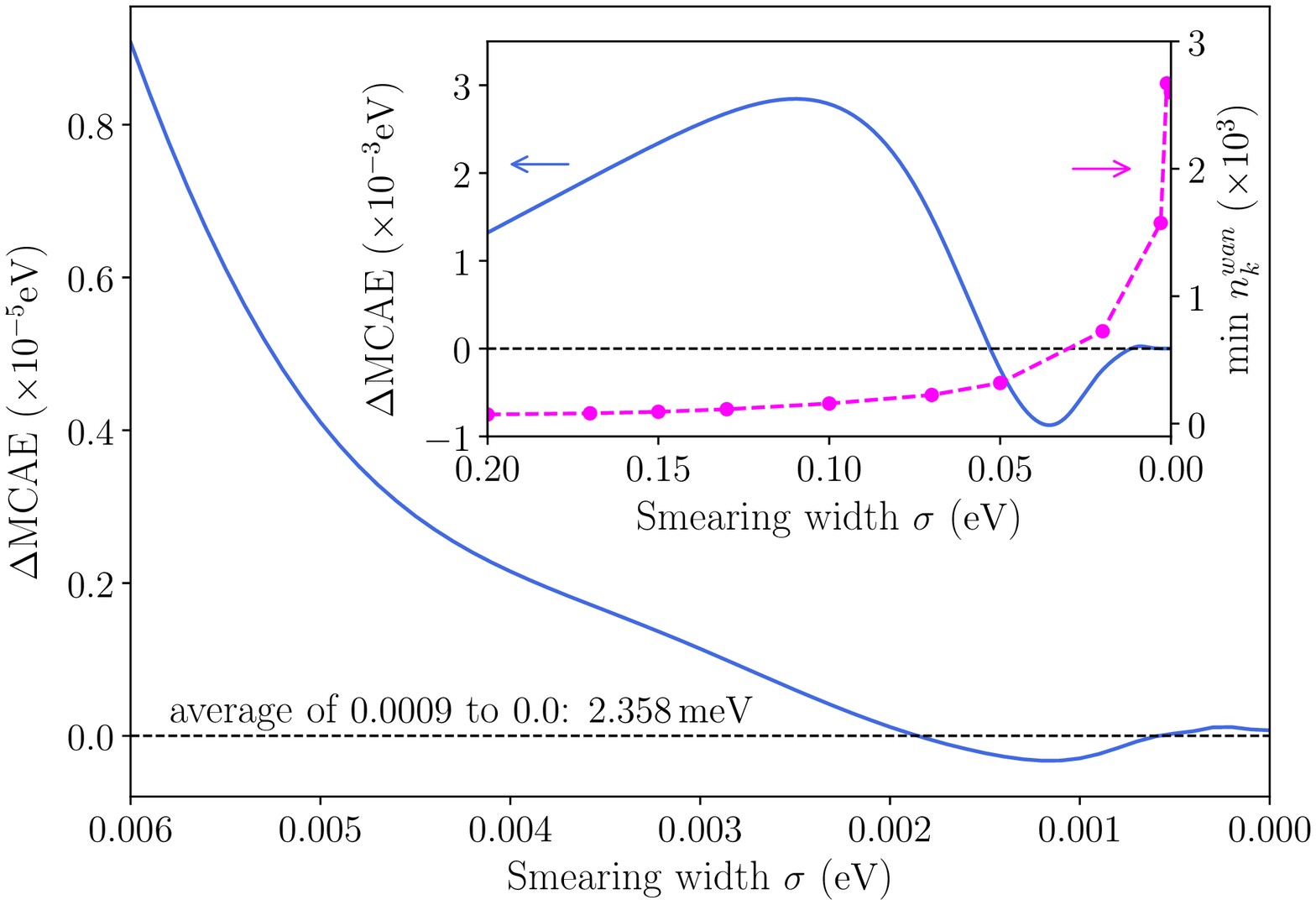}
    \caption{\label{fig:conv_smr} 
        Fe chain MCAE convergence relative to 
        smearing width $\sigma$, defined by 
        Equ. (\ref{equ:Delta_mcae_sigma}). 
        The inset is the overview for $\sigma$ from 
        \SIrange{0.20}{0.00}{eV}. 
        The magenta line in the inset is the minimum number of Wannier 
        interpolation kmesh $n_k^{wan}$ needed for 
        each smearing width $\sigma$. }
\end{figure}

The relation between the minimum number of interpolation kmesh 
$n_k^{wan}$ and the smearing width $\sigma$ is shown in the inset 
of Fig. \ref{fig:conv_smr} and the specific values are tabulated 
in Table \ref{tab:smr_kpt}. When $\sigma$ is in the range of 
\SIrange{0.20}{0.05}{eV}, only less than \num{300} $n_k^{wan}$ 
is needed to converge the MCAE on the order of \SI{2e-6}{eV}. 
However, as indicated by 
the blue line in the inset of Fig. \ref{fig:conv_smr}, the 
smearing contribution to the MCAE,  
$\Delta\text{MCAE}(\sigma;n_k^{wan})$, is on the order of \SI{2e-3}{eV}, leading to unreliable 
results. 
$\Delta\text{MCAE}(\sigma;n_k^{wan})$ is defined as 
\begin{align}
\begin{split}
\label{equ:Delta_mcae_sigma}
& \Delta\text{MCAE}(\sigma;n_k^{wan}) = \text{MCAE}(\sigma;n_k^{wan}) 
- \overline{\text{MCAE}}, 
\\
& \overline{\text{MCAE}} = \frac{1}{10} \sum_{\sigma \in \Sigma} 
\text{MCAE}(\sigma;n_k^{wan}), 
\text{where}~\Sigma = \{0.0001 \times i| 0 \le i \le 9\}, 
\end{split}
\end{align}
where the $n_k^{wan}$ are fixed to 
\num{2600}, and the \num{0.0000} in $\Sigma$ means calculation without smearing.

We find a smearing width of less than \SI{0.003}{eV} is able 
to reduce the smearing contribution to under \SI{2e-6}{eV}, 
and at such a minimal smearing width a kmesh of nearly $1600$ 
should be adopted. To totally exclude the smearing contribution,  
i.e. calculation without smearing, 
a dense kmesh of $2600$ should be used [data listed in 
Table \ref{tab:smr_kpt}]. 

Finally, as the best touchstone for the validity 
of Wannier interpolation, we performed an extremely 
accurate \textit{ab-initio} calculation, with a 
kmesh of $1 \times 1 \times 2600$ and a smearing 
width of \SI{0.0010}{eV}. As shown in Table 
\ref{tab:smr_kpt}, considering the numerical 
accuracy of \SI{0.001}{meV}, 
the Wannier interpolation 
results of rows No.\numrange{9}{11} are exactly 
equal to the dense kmesh \textit{ab-initio} 
result of row No.15.  
The MLWFs used for interpolating MCAEs in rows 
No.\numrange{1}{11} are constructed from the 
\textit{ab-initio} calculation of row No.13.
It is clear from rows No.\numrange{12}{14} 
that the \textit{ab-initio} 
calculation of row No.13 is not converged 
but the constructed MLWFs successfully 
recover the ``true'' MCAE.

\begin{table}
    \caption{\label{tab:smr_kpt} 
    	The Fe chain MCAE and the 
        minimum Wannier interpolation 
        kmesh needed for convergence 
        with respect to  smearing 
        width $\sigma$. The data in this 
        table are plotted in the inset of Fig. \ref{fig:conv_smr}. 
        The MCAEs in rows of 
        No.\numrange{12}{15} are 
        directly calculated by QE, 
        as indicated by QE in 
        the Notes column. While 
        the MCAEs in rows of 
        No.\numrange{1}{11} are calculated by Wannier interpolation, as indicated by Wan in the Notes column. 
    The $\sigma = 0.0000$ in the row No.11 represents 
    calculation without smearing. 
	The MLWFs used for interpolating MCAEs in rows 
	No.\numrange{1}{11} are constructed from the 
	\textit{ab-initio} calculation of row No.13. }
    \begin{center}
        \begin{tabular}{c c c c c c}
            \hline
            No. & $\sigma$ (\si{eV}) & $n_k^{ab}$ & MCAE (\si{meV}) & min $n_k^{wan}$ & Notes \\
            \hline
            1 & \num{0.2000} & 24  &  \num{3.67862}  &   \num{72}  & Wan \\
            2 & \num{0.1700} & 24  &  \num{4.29854}  &   \num{81}  & Wan \\
            3 & \num{0.1500} & 24  &  \num{4.69843}  &   \num{93}  & Wan \\
            4 & \num{0.1300} & 24  &  \num{5.03960}  &   \num{113} & Wan \\
            5 & \num{0.1000} & 24  &  \num{5.14434}  &   \num{158} & Wan \\
            6 & \num{0.0700} & 24  &  \num{3.85764}  &   \num{225} & Wan \\
            7 & \num{0.0500} & 24  &  \num{2.12176}  &   \num{319} & Wan \\
            8 & \num{0.0200} & 24  &  \num{2.11546}  &   \num{725} & Wan \\
            9 & \num{0.0030} & 24  &  \num{2.35876}  &   \num{1575} & Wan \\
            10 & \num{0.0012} & 24  &  \num{2.35821}  &   \num{2670} & Wan \\
            11 & \num{0.0000} & 24  &  \num{2.35739}  &   \num{2598} & Wan \\
            \hline
            12 & \num{0.0272} & 23 & \num{3.26219} & None & QE \\
            13 & \num{0.0272} & 24 & \num{1.76603} & None & QE \\
            14 & \num{0.0272} & 25 & \num{1.23916} & None & QE \\
            \hline
            15 & \num{0.0010} & 2600 & \num{2.35867} & None & QE \\
            \hline
        \end{tabular}
    \end{center}
\end{table}

\subsection{Computational resources}

So far we have demonstrated that our Wannier 
interpolation approach for calculating MCAE is sufficiently 
accurate. To illustrate its effectiveness in relieving the 
computational cost, we directly compare the resources used 
in the different calculations. 

The computational resources and time spent for each calculation 
are shown in Table \ref{tab:resources}. All of the 
\textit{ab-initio} results are calculated on \num{96} CPU cores.
The \textit{ab-initio} calculation on kmesh \num{2600} took 
\SI{9.0}{\hour}, while \textit{ab-initio} calculation on 
kmesh \num{24} took several minutes. Note the most 
time-consuming step in the Wannier interpolation 
method\textemdash the disentanglement and Wannierization 
processes\textemdash can be 
boosted by selecting better initial projection orbitals. 
In the current work, we do not further investigate 
the choice of initial projection orbitals and we 
expect a better choice can greatly reduce the time 
spent on disentanglement and Wannierization. 
The 
biggest advantage of the Wannier interpolation method is that 
once the MLWFs are constructed, interpolations on kmesh of 
arbitrary density can be performed with negligible time. 
Apart from the computational time being saved, the computational 
resources needed for the interpolation is rather small, 
only 1 CPU core is sufficient for interpolation on 
arbitrary kmesh. If simply use 
\begin{equation}
\label{equ:cost}
\text{Cost} = \text{CPU cores} \times \text{Wall time}, 
\end{equation}
as an estimation of computational cost, the direct 
\textit{ab-initio} calculation costs \SI{864}{hour \cdot core}, 
the total cost of the Wannier interpolation method is 
\SI{18}{hour \cdot core}, i.e. we achieved a \num{48} times 
speedup.

\begin{table}
    \caption{\label{tab:resources}Computational resources spent 
        by direct dense kmesh \textit{ab-initio} calculation and 
        by Wannier interpolation method in the MCAE 
        calculations of Fe chain. The cost is defined 
        by Equ. (\ref{equ:cost}). 
    The Time in the third column is the wall time, and  
    	they are the sum of calculations for two magnetization directions. 
    The unit of Cost is hour$\cdot$core and is the 
    sum of all the calculation steps. 
	The Time = 0 in the last row means the time spent was 
	less than a second, negligible.}
    \begin{center}
        \begin{tabular}{l c c c}
            \hline
            \multicolumn{1}{c}{Calculations} & 
            CPU cores & 
            Time & 
            Cost
            \\
            \hline
            direct \textit{ab-initio}, kmesh 2600   &  
            \num{96}    &    
            \SI{9.0}{\hour}   &
            864 
            \\
            \hline
            Wannier interpolation & 
            & 
            &
            \multirow{5}{*}{18}
            \\
            \hspace{3mm} a. coarse \textit{ab-initio}, kmesh 24 &  \num{96} & \SI{5.8}{\minute} \\
            \hspace{3mm} b. projections \& overlaps & \num{96} & \SI{3.8}{\minute} \\
            \hspace{3mm} \makecell[l]{c. disentanglement \& \\ \hspace{3mm} Wannierization} & \num{1} & \SI{2.5}{\hour} \\
            \hspace{3mm} d. interpolation, kmesh 2600 & \num{1} & 0 \\
            \hline
        \end{tabular}
   \end{center}
\end{table}

We comment here that the adaptive refinement of kmesh and 
the adaptive smearing algorithms frequently used in MLWF 
interpolation of other physical properties 
\cite{Yates2007,Marzari2012} are not 
necessary since it is already 
enough convenient to interpolate MCAE 
on arbitrary dense kmesh. For MCAE calculations, smearing 
should be avoided and the main concern is the quality of 
MLWFs. Once high-quality MLWFs are constructed, the 
following interpolation can be easily performed with 
very high accuracy and negligible cost.  

\section{Fe monolayer}
The prototypical calculations on Fe chain have demonstrated 
the effectiveness of Wannier interpolation and we have 
 discussed 
several convergence issues when employing Wannier interpolation to calculate MCAE. In this section, we perform 
calculations on Fe monolayer, which can be directly 
validated against former results in the literature. 
In the next section we will show calculations on 
FeNi alloy, which can be compared with 
many \textit{ab-initio} calculations and experimentally 
measured values.

The MCAE of Fe monolayer was calculated as 
\SI{0.7}{meV} based 
on ultrasoft pseudopotential (USPP) result \cite{Li2013}. 
To make a direct comparison, we used the same structure 
as in Ref.\cite{Li2013}, i.e. (001) Fe monolayer with 
lattice constant \SI{2.85}{\angstrom} and 
\SI{8}{\angstrom} vacuum separation in the z-direction.
The \textit{ab-initio} calculations were performed using 
QE based on PAW-GGA. 
We used a wave function cutoff of \SI{60}{Ry} and 
an electron density 
cutoff of \SI{500}{Ry}. A Methfessel-Paxton 
smearing of width \SI{0.05}{eV} 
was adopted. To accurately determine MCAE, a kmesh of 
$50 \times 50 \times 1$ was used in the SCF calculation. 

\begin{figure}
	\includegraphics[width=\columnwidth]{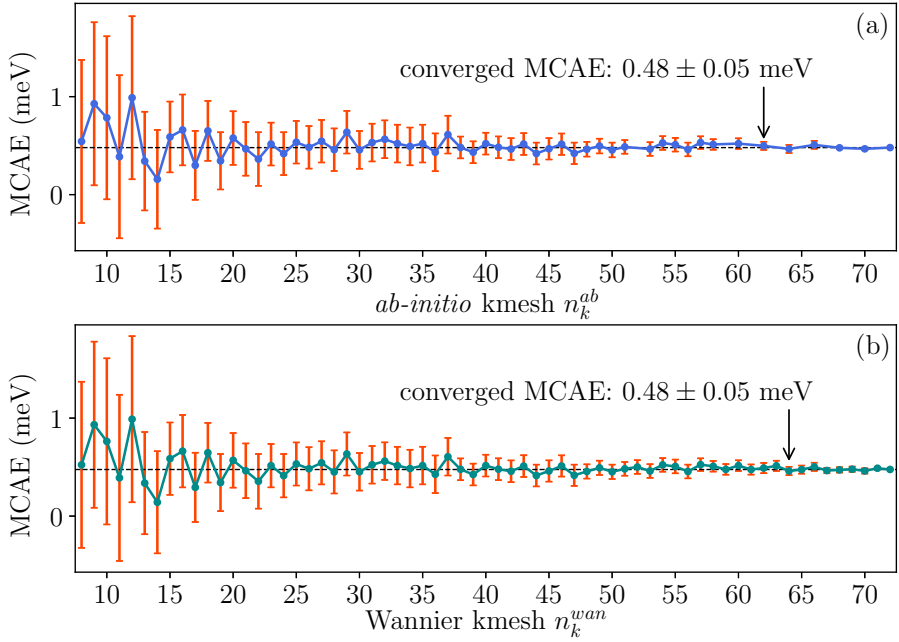}
	\caption{\label{fig:femonolayer} 
		Fe monolayer MCAE convergence relative to 
		(a) \textit{ab-initio} kmesh 
		$n_k^{ab}$, and 
		(b) Wannier interpolation kmesh 
		$n_k^{wan}$. 
		The MLWFs were constructed from 
		\textit{ab-initio} kmesh 
	$11 \times 11 \times 1$, and it is clear 
	from (a) that this kmesh is far from 
	convergence if evaluating MCAE by direct 
	\textit{ab-initio} calculation. In fact, 
	a kmesh of $62 \times 62 \times 1$ is needed 
	for the \textit{ab-initio} calculation to 
	converge. The error bars are the 
	convergence indicators obeying the definition 
	of Equ.(\ref{equ:delta_f}). 
	To save computational time, we selectively 
	calculated some of the kmeshes 
	when $n_k^{ab} > 60$ in (a). }
\end{figure}

As is shown in Fig.\ref{fig:femonolayer}(a), for a direct 
\textit{ab-initio} calculation,  
a kmesh higher than $62 \times 62 \times 1$ is needed 
to converge MCAE under \SI{0.05}{meV}, and the 
converged value of $0.48\pm 0.05$\si{meV} 
(magnetic moment perpendicular to the plane) 
is similar 
to that of $0.70$\si{meV} in Ref.\cite{Li2013}. 
The \SI{0.20}{meV} difference between our results and 
the Ref.\cite{Li2013} may be caused by different 
type of pseudopotentials. 

Then we 
constructed MLWF from coarse kmesh \textit{ab-initio} 
calculations and it was found that a kmesh of 
dimension $11 \times 11 \times 1$ was enough for the 
convergence of MCAE when using Wannier interpolation. 
The results of the convergence tests are compiled 
in Table \ref{tab:femonolayer}. 
Apparently, from Fig.\ref{fig:femonolayer}(a) we can conclude that \textit{ab-initio} calculations of kmesh around $11 \times 11 \times 1$ are far from convergence, but from 
Table \ref{tab:femonolayer} rows No.\numrange{4}{6}, 
the $11 \times 11 \times 1$ kmesh 
\textit{ab-initio} calculation is sufficient for the 
Wannier interpolation of MCAE. With the help of MLWF, 
equivalent accuracy MCAE value can be extracted from a 
calculation with only 3\% the number of 
$k$-points needed for the converged \textit{ab-initio} calculation. The accuracy of Wannier interpolation 
is even more convincing when comparing Fig.\ref{fig:femonolayer}(a) and (b). Two curves are 
nearly identical, i.e. the MCAE is faithfully recovered 
for each kmesh on the horizontal axis. 

As has been shown in Fig.\ref{fig:conv_smr} and 
Table \ref{tab:smr_kpt}, smearing can be used 
to inhibit MCAE fluctuation and greatly reduce 
the density of kmesh needed for the convergence. 
However, the fictitious smearing will 
give rise to deviations around converged MCAE. 
In such circumstance, the Wannier interpolation 
is very helpful because we can cheaply increase 
the density of kmesh without resorting to smearing. 
We calculated the MCAE of Fe monolayer 
with different smearing width [Table 
\ref{tab:femonolayer} rows No.\numrange{7}{9}] and 
it was found that kmesh should increase to 
more than $86 \times 86 \times 1$ to converge 
the MCAE. Since we have already used a small 
smearing width of \SI{0.05}{eV} in the 
\textit{ab-initio} calculation, the MCAE hardly 
changes when reducing the smearing width. One 
noticeable irregularity is that the kmesh 
for convergence decreases when smearing width 
decreases. We speculate that this is caused by 
small numerical noises rather than a general 
trend. 

In all, for the MCAE on the order of sub \si{meV}, 
the Wannier interpolation successfully 
recovers high accuracy \textit{ab-initio} 
results from coarse kmesh \textit{ab-initio} calculations.

\begin{table}
	\caption{\label{tab:femonolayer}The convergence of 
		Fe monolayer MCAE 
		with respect to the \textit{ab-initio} 
		kmesh $n_k^{ab}$, 
		Wannier interpolation kmesh $n_k^{wan}$ 
		and smearing width $\sigma$.
		The MCAEs in rows of 
	No.\numrange{1}{3} are directly calculated by QE, 
	while the MCAEs in rows of 
	No.\numrange{4}{9} are calculated by Wannier interpolation. The No.10 row is the result of 
	USPP-GGA calculation from Ref.\cite{Li2013}. 
    The number $n_k$ in columns 2 and 3 stands for 
	kmesh $n_k \times n_k \times 1$. }
\begin{center}
	\begin{tabular}{c c c c c c}
		\hline
		No. & $n_k^{ab}$ & $n_k^{wan}$ & $\sigma$ (\si{eV}) & MCAE (\si{meV}) & Notes \\
		\hline
		1 & 10 & None  & 0.05 & $0.78$ & QE\\
		2 & 11 & None  & 0.05 & $0.39$ & QE\\
		3 & $\ge$62 & None  & 0.05 & $0.48 \pm 0.05$ & QE\\
		\hline
		4 & 10 & $\ge$74 & 0.05 & $0.67 \pm 0.05$ & Wan\\
		5 & 11 & $\ge$64 & 0.05 & $0.48 \pm 0.05$ & Wan\\
		6 & 12 & $\ge$64 & 0.05 & $0.47 \pm 0.05$ & Wan\\
		\hline
		7 & 11 & $\ge$109 & 0.02 & $0.50 \pm 0.05$ & Wan\\
		8 & 11 & $\ge$99 & 0.01 & $0.50 \pm 0.05$ & Wan\\
		9 & 11 & $\ge$86 & 0.00 & $0.50 \pm 0.05$ & Wan\\
		\hline
		10 & 40 & None & 0.05 & $0.70$ & USPP\cite{Li2013}\\
		\hline
	\end{tabular}
\end{center}
\end{table}

\section{FeNi alloy}
MCAE tends to be larger when symmetry is lower, and 
the MCAE of surfaces and interfaces are on the order 
of \si{meV}. Our calculations on Fe chain and Fe monolayer 
have validated that Wannier interpolation is capable of 
efficiently reducing the computational costs for MCAE 
in the region of \si{meV}. The calculation of 
bulk MCAE is much harder since their MCAE are usually 
on the order of \si{\mu eV}. 
What is even worse is that in 3 dimensions we need 
to sample the BZ along 3 directions, this means 
the number of $k$-points will increase cubically 
when we densify the kmesh for the convergence of 
MCAE. 
To further demonstrate the usefulness of Wannier 
interpolation, we performed calculations on 
FeNi alloy since a practical comparison between 
our Wannier interpolation results and other \textit{ab-initio} calculations as well 
as experimental results can be made. 
Also, as a demonstration of the versatility of 
Wannier interpolation,  
we use {\sc{GPAW}}\cite{PhysRevB.71.035109,Enkovaara2010Jun} as the \textit{ab-initio} calculator 
in this subsection, 
since the construction of MLWF is irrelevant 
with the kind of the underlying \textit{ab-initio} code 
and the Wannier 
interpolation approach is a post-processing tool 
capable of cooperating with different \textit{ab-initio} calculator seamlessly. 

Based on \textit{ab-initio} calculations and 
experimental measurements, the MCAE of FeNi alloy is 
in the range of \SIrange{0.48}{0.77}{MJ/m^3} 
\cite{Mizuguchi2011Jul,Edstrom2014Jul,
	Miura2013Feb,Shima2007Mar}. 
We used the same structure as 
in Ref.\cite{Miura2013Feb,Edstrom2014Jul}, 
i.e. $L1_0$ FeNi with in-plane lattice 
constant $a=3.56$\si{\angstrom} and out-of-plane 
lattice constant $c=3.58$\si{\angstrom}. 
The calculated cell is \SI{45}{\degree} rotated relative 
to the conventional cell and the in-plane lattice 
constant is $a/\sqrt{2}$. 
The \textit{ab-initio} calculations were performed using 
GPAW based on PAW-GGA with \SI{600}{eV} 
wave function cutoff and a Fermi-Dirac  
smearing of width \SI{0.1}{eV}. 
A kmesh of $25 \times 25 \times 17$ was used in the SCF calculation and the convergence criterion was set 
as \SI{1e-6}{eV}.

\begin{figure}
	\includegraphics[width=\columnwidth]{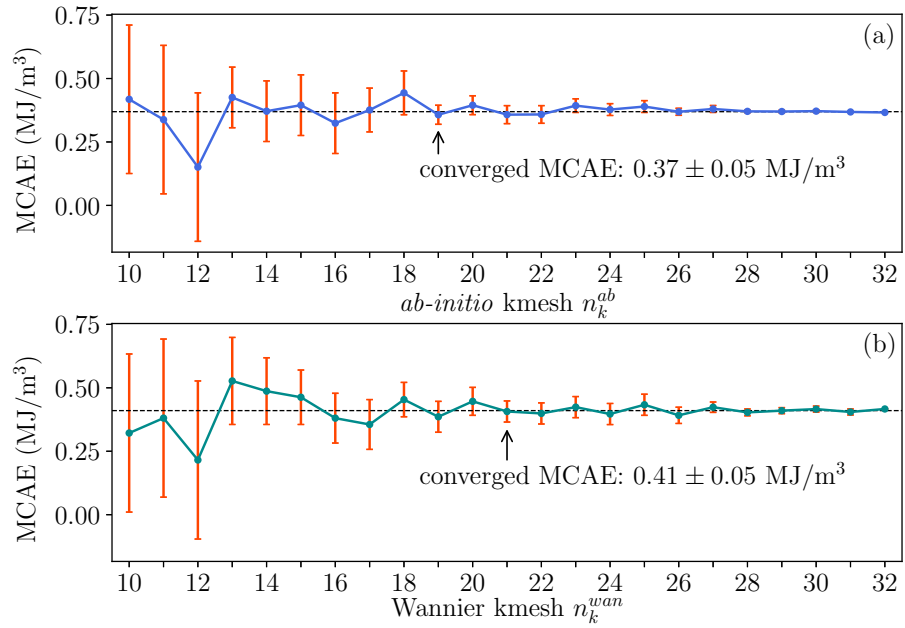}
	\caption{\label{fig:feni} 
		FeNi alloy MCAE convergence relative to 
		(a) \textit{ab-initio} kmesh 
		$n_k^{ab}$, and 
		(b) Wannier interpolation kmesh 
		$n_k^{wan}$. 
		The MLWFs were constructed from 
		\textit{ab-initio} kmesh 
		$12 \times 12 \times 8$, and it is clear 
		from (a) that this kmesh is far from 
		convergence if evaluating MCAE by direct 
		\textit{ab-initio} calculation. In fact, 
		a kmesh of $19 \times 19 \times 13$ is needed 
		for the \textit{ab-initio} calculation to 
		converge. The error bars are the 
		convergence indicators obeying the definition 
		of Equ.(\ref{equ:delta_f}). }
\end{figure}

For the \textit{ab-initio} calculation, a kmesh 
of dimension $19 \times 19 \times 13$ is needed 
to converge MCAE [Fig.\ref{fig:feni} and Table 
\ref{tab:feni} rows No.\numrange{1}{3}]. The 
converged MCAE is \SI{52}{\mu eV} 
(magnetic moment along [001] axis), i.e. 
\SI{0.37}{MJ/m^3}, a bit smaller than 
the value in the literature around 
\SI{0.5}{MJ/m^3} 
\cite{Mizuguchi2011Jul,Edstrom2014Jul,
	Miura2013Feb,Shima2007Mar}.
Next, we construct MLWF based on coarse kmesh 
\textit{ab-initio} calculations. It was 
found that kmesh of dimension $12 \times 12 \times 8$ 
is sufficient for constructing MLWFs and 
a kmesh of $21 \times 21 \times 21$ is enough for 
Wannier interpolation. 
Due to the limitations of GPAW-Wannier90 interface, 
the quality of the constructed MLWFs was not 
so good as that of QE. Nevertheless, 
the Wannier interpolated MCAE of 
$0.41\pm 0.05$\si{MJ/m^3} [Table \ref{tab:feni} 
row No.6], still successfully 
recovered high density kmesh \textit{ab-initio} 
result of $0.37\pm 0.05$\si{MJ/m^3} 
[Table \ref{tab:feni} row No.3]. 
In this case, MLWFs enable us to get equivalent 
accuracy MCAE result from coarse kmesh of 
\SI{25}{\percent} the number of $k$-points relative 
to the dense kmesh \textit{ab-initio} calculation. 
We expect that by improving the GPAW-Wannier90 
interface, the density of the 
$12 \times 12 \times 8$ kmesh for constructing 
MLWFs can be further reduced, thus higher 
efficiency when using Wannier interpolation. 

In the \textit{ab-initio} calculations we used 
a large smearing width of \SI{0.1}{eV}, next 
we utilize Wannier interpolation to recover the 
MCAE without smearing. Upon decreasing smearing 
width to 0, a Wannier interpolation kmesh of 
dimension $67 \times 67 \times 67$ is needed 
to converge MCAE [Table \ref{tab:feni} row No.7] 
and the result of 
$0.52\pm 0.05$\si{MJ/m^3} is quite similar 
to the literature. A $67 \times 67 \times 67$ 
kmesh means that the number of $k$-points is 
on the order of \num{300000}, such a large number 
of $k$-points poses serious challenges to 
\textit{ab-initio} calculations. However, since 
in MLWF space the number of WFs is on the order of 
\numrange{10}{100}, the diagonalization of the 
Hamiltonian matrix is much cheaper thus the 
eigenvalues are efficiently computed. 
The Wannier interpolation is very helpful 
when it requires an extremely dense kmesh, and 
the case of FeNi alloy has proven that 
Wannier interpolation is capable of computing 
even bulk magnetocrystalline anisotropy energy.

\begin{table}
	\caption{\label{tab:feni}The convergence of FeNi alloy MCAE 
		with respect to the \textit{ab-initio} 
		kmesh $n_k^{ab}$, 
		Wannier interpolation kmesh $n_k^{wan}$ 
		and smearing width $\sigma$.
		The MCAEs in rows of 
		No.\numrange{1}{3} are directly calculated by GPAW, 
		while the MCAEs in rows of 
		No.\numrange{4}{7} are calculated by Wannier interpolation. The No.\numrange{8}{10} rows are the results of VASP\cite{Miura2013Feb}, WIEN2K\cite{Edstrom2014Jul} 
		and SPR-KKR\cite{Edstrom2014Jul} calculations 
		respectively. The No.11 and 12 rows are the results of experiments\cite{Shima2007Mar,Mizuguchi2011Jul}. 
		The number $n_k$ in columns 2 and 3 stands for 
		kmesh $n_k \times n_k \times int(\frac{a}{c}n_k)$ 
		where $a$ and $c$ are the in-plane and out-of-plane 
		lattice constants, $int(x)$ means truncating real 
		number $x$ into its integer part, i.e. 
		$n_k \times n_k \times int(\frac{a}{c}n_k)$ means 
		a uniformly distributed kmesh. The $n_k^{ab}$ 
		in rows No.\numrange{8}{10} are the total number 
		of $k$-points in the BZ.}
	\begin{center}
	\begin{tabular}{c c c c c c}
		\hline
		No. & $n_k^{ab}$ & $n_k^{wan}$ & $\sigma$ (\si{eV}) & MCAE (\si{MJ/m^3}) & Notes \\
		\hline
		1 & 11 & None  & 0.1 & $0.34$ & GPAW\\
		2 & 12 & None  & 0.1 & $0.15$  & GPAW\\
		3 & $\ge$19 & None  & 0.1 & $0.37 \pm 0.05$ & GPAW\\
		\hline
		4 & 10 & $\ge$21 & 0.1 & $0.43 \pm 0.05$ & Wan\\
		5 & 11 & $\ge$22 & 0.1 & $0.54 \pm 0.05$ & Wan\\
		6 & 12 & $\ge$21 & 0.1 & $0.41 \pm 0.05$ & Wan\\
		\hline
		7 & 12 & $\ge$67 & 0.0 & $0.52 \pm 0.05$ & Wan\\
		\hline
		8 & 19800 &   &  & $0.56$ & VASP\cite{Miura2013Feb}\\
		9 & 20000 &  &  & $0.48$ & WIEN2K\cite{Edstrom2014Jul}\\
		10 & 160000 &   &  & $0.77$ & SPR-KKR\cite{Edstrom2014Jul}\\
		\hline
		11 &   &   &   & $0.63$ & Exp.\cite{Shima2007Mar}\\
		12 &   &   &   & $0.58$ & Exp.\cite{Mizuguchi2011Jul}\\
		\hline
	\end{tabular}
\end{center}
\end{table}

\section{Conclusions}

We combine the Wannier interpolation of eigenvalues and 
the force theorem of magnetocrystalline anisotropy energy 
together to develop an efficient and accurate method for 
calculating MCAE. First, coarse kmesh \textit{ab-initio} 
calculations are performed for different magnetization 
directions. Second, maximally localized Wannier functions 
are constructed based on the overlap matrices 
and projection matrices obtained from the first step. 
Third, MCAE is calculated based on FT and 
Wannier interpolation of eigenvalues on a dense kmesh. 
This Wannier interpolation method serves as 
a post-processing step for economically calculating MCAE 
other than brute-force \textit{ab-initio} calculation. 
The ultimate accuracy of the calculated MCAE is determined 
by the underlying \textit{ab-initio} code (the choice of 
exchange-correlation functionals, type of basis sets, etc.), 
since it is the \textit{ab-initio} code that produces the 
Bloch functions from which the MLWFs are constructed.  
Nevertheless, the Wannier interpolation is independent of the 
choice of the underlying code, since the only quantities 
needed from the code are the overlap and projection matrices 
which can be obtained merely from Bloch functions.  

First, we take a Fe chain as an example and demonstrate that 
our Wannier interpolation approach for MCAE achieves 
a \num{48} times speedup compared to the direct dense kmesh 
\textit{ab-initio} calculation, and in this example 
we discuss three critical factors for successfully 
getting high accuracy MCAE: the coarse \textit{ab-initio} 
kmesh for constructing MLWFs, the dense 
Wannier interpolation kmesh and the smearing width. 
It is found that the most important issue is 
the quality of MLWFs, and we need a sufficiently dense 
kmesh to reach a reliable MCAE, 
but this kmesh is still much sparser than 
direct \textit{ab-initio} calculation. 
The Wannier interpolation kmesh can be easily 
increased and smearing width can be reduced to 0. 
Then, the Fe monolayer and the FeNi alloy example 
demonstrate that our Wannier interpolation approach 
is capable of calculating MCAE on the order of 
\si{meV} to \si{\mu eV}, greatly mitigating the 
computational burden on calculating 
interface and bulk MCAE. 
In summary, 
to get the most accurate MCAE, no smearing should be 
adopted and the coarse kmesh \textit{ab-initio} calculation 
should be fully tested. The convergence relative to the 
Wannier interpolation kmesh can be easily achieved. 

The cooperations of QE-Wannier90 and 
GPAW-Wannier90 demonstrate that the Wannier 
interpolation approach can work with different 
\textit{ab-initio} calculators seamlessly. 
We expect that the Wannier interpolation 
approach for MCAE can be more useful 
when the quality of the interfaces between 
\textit{ab-initio} calculators and Wannier90 
are improved. 
This Wannier interpolation 
method reduces the computational cost significantly 
and maintains high accuracy simultaneously. 
Besides, it makes it possible for MCAE calculations which are 
hard to converge or unfeasible due to the computational cost, 
as well as high-throughput calculations to identify material 
candidates for spintronics devices.

\section*{Acknowledgments}
This work was supported by the National Natural Science 
Foundation of China [grant numbers 61627813, 61571023]; 
the International Collaboration Project [grant number B16001]; 
and the National Key Technology Program of China [grant number 2017ZX01032101]. 


\bibliography{refs}

\end{document}